\documentstyle[times,graphics,epsfig,amsmath,amssymb,natbib,referee]{mn2e}
\newcommand{\be}{\begin{equation}}
\newcommand{\e}{\end{equation}}
\newcommand{\bear}{\begin{eqnarray}}
\newcommand{\ear}{\end{eqnarray}}

\def\aj{AJ}
\def\apj{ApJ}

\def\mnras{MNRAS}

\def\aap{A\&A}
\def\apjs{ApJS}

\begin{document}
\title[Exploring the Cosmic Web in SDSS DR7]{Exploring the Cosmic Web
  in the Sloan Digital Sky Survey Data Release Seven using the Local
  Dimension.}

\author[Sarkar P.,Pandey B., Bharadwaj, S.]  {Prakash
  Sarkar$^{1}$\thanks{E-mail:sarkar@iucaa.ernet.in}, Biswajit Pandey
  $^{2,4}$\thanks{Email:biswa@mpa-garching.mpg.de}, Somnath
  Bharadwaj$^{3}$\thanks{Email:somnath@phy.iitkgp.ernet.in}
  \\$^1$IUCAA, Pune University Campus, Post Bag 4, Ganeshkhind, Pune
  411 007, India.\\$^2$Max-Planck Institute for Astrophysics,
  Karl-Schwarszchild Str. 1, D85748, Garching, Germany
  \\$^{3}$Department of Physics and Meteorology \& Centre for
  Theoretical Studies , IIT Kharagpur, 721 302 , India
  \\$^{4}$Department of Physics, Visva-Bharati University,
  Santiniketan, Birbhum, 731235, India \\ }

\maketitle

\date{\today}

\begin{abstract}
It is possible to visualize the Cosmic Web as an interconnected
network of one-dimensional filaments, two-dimensional sheets and
three-dimensional volume-filling structures which we refer to as
clusters.We have used the Local Dimension $D$, which takes values
$D=1, 2$ and $3$ for filaments, sheets and clusters, respectively, to
analyse the Cosmic Web in a three-dimensional volume-limited galaxy
sample from the Sloan Digital Sky Survey Data Release 7. The analysis
was carried out separately using three different ranges of
length-scales: 0.5-5, 1-10 and 5-50 $h^{-1} {\rm Mpc}$. We find that
there is a progressive increase in the $D$ values as we move to larger
length-scales. At the smallest length-scale, the galaxies
predominantly reside in filaments and sheets. There is a shift from
filaments to sheets and clusters at larger scales. Filaments are
completely absent at the largest length-scale (5–50 $h^{-1} {\rm
  Mpc}$). Considering the effect of the density environment on the
Cosmic Web, we find that the filaments preferentially inhabit regions
with a lower density environment as compared to sheets and clusters
which prefer relatively higher density environments. A similar
length-scale dependence and environment dependence was also found in a
galaxy sample drawn from the Millennium Simulation which was analysed
in exactly the same way as the actual data.

\end{abstract}


\section{Introduction}
Galaxy redshift surveys provide us with a picture of the large
scale structures in the present day universe. All the major galaxy
redshift surveys like the Center for Astrophysics (CfA) Survey
\citep{gel}, the Las Campanas Redshift Survey (LCRS)
\citep{shect}, the Two-Degree Field Galaxy Redshift Survey
(2dFGRS) (\citealt{colless}) and the Sloan Digital Sky Survey
(SDSS) (\citealt{york}) clearly show that the galaxies are
distributed in a complex interconnected network of filaments,
sheets and clusters encircling nearly empty voids. This
interconnected network is often referred to as the ``Cosmic Web''
\citep{bond96}. Quantifying the Cosmic Web and understanding it's
origin is one of the most interesting and challenging issues in
cosmology.

A wide variety of statistical measures have so far been employed to
quantify the Cosmic Web. The void probability function
\citep{White1979}, the percolation analysis \citep{shandarin1983} and
the genus curve \citep{Gott1} are some of the earliest statistics
introduced to quantify the topology of the galaxy distribution.  The
Minkowski functionals \citep{mecke1994} provide a global
characterization of structures. Ratios of the Minkowski functionals
can be used to define the `Shapefinders' which are a set of shape
diagnostics for both simple and topologically complex objects
(\citealt{Sahni1998}). \citet{bharadwaj2000} have introduced a two
dimensional (2D) version of Shapefinders. There are quite a few
different techniques that have been introduced to identify voids in
the galaxy distribution \citep{hagai,hoylea,platen,neyrinck1}.
\citet{stoi} have proposed a three dimensional object point process to
delineate filaments in the large scale structures. \citet{colombi}
have proposed a Hessian based statistics to study the topology of
excursion sets at the percolation threshold.  The Smoothed Hessian
Major Axis Filament Finder (SHMAFF) (\citealt{bondstrausscen}) has
been introduced to identify filaments, sheets and clusters in the
galaxy distribution. \citet{arag} have introduced a multiscale
morphology filter to automatically segment the cosmic structures into
its basic components namely clusters, filaments and
walls. \citet{sous} have proposed a skeleton formalism to quantify
filamentary structures in a three dimensional density field.
\citet{aragon} have introduced the Spine of the Cosmic Web which
provides a complete framework for the identification of different
morphological components of the Cosmic Web. Each of the techniques
mentioned above quantifies one or atmost a few aspects of the complex
network referred to as the Cosmic Web. A comprehensive quantification
of the Cosmic Web is still forthcoming, leaving considerable scope for
work in this direction.

The early galaxy redshift surveys like the LCRS probed thin, nearly
two dimensional (2D) slices through the universe.  Initial
investigations \citep{bharadwaj2000} which used the 2D Shapefinders
show the galaxy distribution in the LCRS to have excess filamentarity
in comparison to a random distribution of
points. \citet{bharadwaj2004} show the filamentarity in the LCRS to be
statistically significant up to a length-scales of $70-80 \, h^{-1} \,
{\rm Mpc}$ but not beyond. Longer filaments, though present in the
data, were not found to be statistically significant, and were
possibly the outcome of chance alignments of shorter filaments.
\citet{bharadwajpandey2004} have used N-body simulations to show that
the filamentarity in the LCRS is consistent with the LCDM model with a
mild bias.  Subsequent work \citep{pandey} performed a 2D analysis of
thin slices from the SDSS. This confirmed the results obtained earlier
using the LCRS. It was also found that the distribution of brighter
galaxies has a lower connectivity and filamentarity as compared to the
fainter ones.  The filamentarity was also found to depend on other
galaxy properties like colour and morphology \citep{pandey2}, and the
star formation rate \citep{pandey4}.  \citet{pandey3} have computed
the filamentarity for galaxy samples in different luminosity bins and
compared these to the filamentarity in simulated galaxy samples with
different values of the linear bias parameter to obtain a
luminosity-bias relation. A recent 2D analysis of the luminous red
galaxies in the SDSS shows the filamentarity to be statistically
significant to length-scales as large as $100$ to $130 \, h^{-1} \,
{\rm Mpc}$ \citep{pandey5}, which are considerably larger than those
found earlier in the LCRS and the SDSS main galaxy sample.

Though the Cosmic Web is relatively easy to visualize and analyze in
2D, this has several limitations which can only be overcome in a three
dimensional (3D) analysis.  For example, a 2D filament could actually
be a section through a 3D
sheet. \citet{2009MNRAS.394L..66S}(henceforth Paper I) have proposed
the Local Dimension as a 3D statistics for analysing the Cosmic Web.
This can, in principle, be used to classify different structural
elements along the Cosmic Web as filaments, sheets and clusters. Tests
with cosmological N-body simulations (Paper I) show that the
structures identified a filaments by the Local Dimension match quite
well with the visual appearance. The structures identified as sheets,
however, could not be visually identified. This was attributed to the
fact that the visual appearance is determined by the most dominant
structures in the field which usually are the filament, whereas the
sheets which are relatively diffuse structures are not visually
identified. Paper I also showed that the Local Dimension could also be
used to address a variety of issues like determining the relative
fraction of galaxies in filaments, sheets and clusters respectively.

In the present work we have used the Local Dimension to analyse the
patterns in the galaxy distribution in the Sloan Digital Sky Survey
Data Release Seven (SDSS DR7).  We have also carried out a similar
analysis on a galaxy catalogue from the semi-analytic model of galaxy
formation implemented in the Millennium Simulation \citep{springel},
and compared the results from the actual data with those from the
simulation.  The Local Dimension classifies the different structural
elements along the Cosmic Web as filaments, sheets and clusters. We
have used this to study the relative fraction of galaxies in these
three different kinds of structures. Further, we also study how these
three different kinds of structures are distributed with reference to
varying density environments. For example, it is possible that the
filaments are preferentially distributed in high density environments
relative to the low density environments. It is well accepted that the
Cosmic Web is a complex network whose morphology and connectivity will
depend on the length-scale at which the analysis is carried out.  The
galaxy distribution is expected to approach homogeneity at large
length scales, and it is known \citep{sarkar} that the SDSS DR6 is
consistent with homogeneity at length-scales beyond $60-70 \, h^{-1}
\, {\rm Mpc}$.  Given these considerations, we have carried out the
entire analysis for three different ranges of length-scales namely 0.5
to 5 $h^{-1} {\rm Mpc}$, 1 to 10 $h^{-1} {\rm Mpc}$ and 5 to 50
$h^{-1} {\rm Mpc}$.

An brief outline of the paper follows.  In Section 2 we describe the
Local Dimension which is the method that we have adopted to analyse
the Cosmic Web  and in Section 3 we discussed the data and the method
of analysis while Section 4 contains the results and conclusions.

\section{The Local Dimension}
The Cosmic Web may be thought of as an interconnected network of
different structural elements. Any particular structural element may
be classified as being either a cluster, a filament or a sheet.  The
Local Dimension, introduced in Paper I, is a simple yet effective
method to quantify the shape of individual structural elements along
the Cosmic Web.  The entire analysis is based on the following
argument. Consider, for example, a galaxy ``G'' located in a filament
which is , by definition, a one dimensional structure.  We expect
$N(<R)$, the number of other galaxies within a sphere of comoving
radius $R$ centered on `G', to scale as $N(<R) = A R$. Similarly, we
expect a scaling $N(<R) = A R^D$ with $D=2$ and $3$ if G were located
in a sheet and a cluster respectively.  The exponent $D$ quantifies
the dimension of the galaxy distribution in the neighbourhood of G,
and hence we refer to it as the Local Dimension.  In principle the
Local Dimension provides a technique to determine whether a galaxy is
located in a filament, a sheet or a cluster.

Consider, for example, a part of the Cosmic Web where there is a
filament connected to a sheet (Figure 1 of Paper 1).  We expect $D=1$
and $2$ for the galaxies located in the filament and the sheet
respectively.  We also expect that the galaxies located near the
junction of the two structures will not exhibit a well defined scaling
behaviour, and it will therefore not be possible to determine the
Local Dimension $D$ for these galaxies.  In a typical situation where
the Cosmic Web is a complex, interconnected network of different kinds
of structural elements only a fraction of all the galaxies in the
entire survey will have a definite value of the Local Dimension $D$.

We have, until now, not taken into account the fact that galaxies are
discrete objects with finite, non-zero intergalactic separations.  It
is thus necessary to consider a finite range of comoving length scales
$R \le R_2$ in order to determine if there is a scaling behaviour
$N(<R) = A R^D$ and thereby estimate the Local Dimension $D$. Contrast
this with a continuum where it is possible to study the scaling
behaviour in the limit $R \rightarrow 0$ and it is not necessary to
refer to a finite range of length-scales.  Further, the Cosmic Web is
expected to look different when viewed at different length-scales.  We
also expect the Cosmic Web to approach a homogeneous network beyond
length-scales of $60-70 \, h^{-1} \, {\rm Mpc}$ \citep{sarkar} where
the galaxy distribution is known to approach homogeneity.  It is thus
interesting and useful to separately analyze the scaling behaviour
across different ranges of length-scale.  Based on this we also
introduce a lower length-scale $R_1$ and study the scaling behaviour
over different ranges of length-scales $R_1 \le R \le R_2$.

The values of Local Dimensions once determined, by fitting power law
over the length-scales $R_1 \le R \le R_2$, can be put to use to
address a variety of issues. First, we can identify individual
structures like filaments or sheets.  The relative abundance of
different $D$ values allows us to estimate the fraction of galaxies
that reside in sheets, filaments and clusters respectively.  The
spatial distribution of the $D$ values allowed us to study how the
different kinds of structural elements are interconnected or ``woven
into the Cosmic Web''.

\section{Data and Method of Analysis}

\subsection{SDSS DR7 Data}

Our present analysis is based on galaxy redshift data from the SDSS
DR7 \citep{abazajian09}.  The SDSS DR7 includes $11,663$ square degrees of
imaging and $9380$ square degrees of spectroscopy with $930,000$
galaxy redshifts.  For the present work we have used the Main Galaxy
Sample for which the target selection algorithm is detailed in
\citet{strauss}. The Main Galaxy Sample comprises of galaxies brighter
than a limiting r band Petrosian magnitude $17.77$.  The data was
downloaded from the Catalog Archive Server (CAS) of SDSS DR7 using a
Structured Query Language (SQL) search. We have identified a
contiguous region in the Northern Galactic Cap which spans
$-40^\circ<\lambda<33^\circ$ and $-30^\circ<\eta<30^\circ$ where
$\lambda$ and $\eta$ are survey co-ordinates defined in
\citet{stout}. A volume limited galaxy subsample was constructed in
this region by restricting the extinction corrected Petrosian r band
apparent magnitude to the range $14.5 \leq m_r \leq 17.77$ and
restricting the absolute magnitude to the range $-19 \leq M_r \leq
-20.5$. This gives us $47705$ galaxies in the redshift range $0.035
\leq z \leq 0.076 $ which corresponds to the comoving radial
distance range $104 \, h^{-1} \le r \le 223 \, h^{-1}$ Mpc.  At the
redshift $z=0.035$, the comoving volume corresponding to our
sub-sample subtends $124 \, h^{-1} {\rm Mpc} \, \times 104 \, h^{-1}
{\rm Mpc}$ along the transverse directions.

\subsection{Millennium Data}

The Millennium Simulation \citep{springel} is a large cosmological
N-body simulation which traced $2160^3$ particles from redshift $127$
to the present in a periodic comoving box of side $500 \, h^{-1} {\rm
  Mpc}$. We have used a semi-analytic galaxy catalogue generated by
\citet{guo11} who updated the previously available galaxy formation
models (\citealt{springel, croton, delucia}) with improved versions
and implemented galaxy formation models on the Millennium Simulation.
Semi-analytic models are simplified simulations of the formation and
evolution of galaxies in a hierarchical clustering scenario
incorporating all relevant physics of galaxy formation processes. The
spectra and magnitude of the model galaxies were computed using
population synthesis models of \citet{brujual}. Using the peculiar
velocities, we map the galaxies to redshift space and then identify a
region having the same geometry as our actual data.  Applying the same
magnitude cuts as those used for the actual data, we have extracted
the same number of galaxies as in our final SDSS data and used this in
our subsequent analysis.

\subsection{Method of analysis}
We have determined $N(<R)$ at several $R$ values for each galaxy in
our sample.  For each galaxy, its distance from the survey boundary
sets the largest value of $R$ for which it is possible to estimate
$N(<R)$.  We have assigned the Poisson error $\Delta N(<R) =
\sqrt{N(<R)}$ to each measured value of $N(<R)$.  A $\chi^{2}$
minimisation procedure was used to determine the best fit power law
$N(<R)=A R^{D}$ to the $N(<R)$ measured for each galaxy.  The value of
$D$ is accepted as the Local Dimension corresponding to the particular
galaxy if the chi-square per degree of freedom of the power law fit
satisfies, $\chi^2/\nu \le 1$. The power law fit is rejected for
larger values of $\chi^2/\nu$, and the Local Dimension is undetermined
for these galaxies.  The fitting procedure was restricted to values
within the range $R_1 \le R \le R_2$ in length-scale.

We have carried out the analysis for three different ranges of
length-scales, each covering a decade.  Values of $D$ were determined
separately across the length-scales 0.5 to 5 $h^{-1} {\rm Mpc}$, 1 to
10 $h^{-1} {\rm Mpc}$ and 5 to 50 $h^{-1} {\rm Mpc}$.  It was possible
to determine a definite value of $D$ for 3484, 9082 and 288 galaxies
at the three respective ranges of length-scale mentioned above.

\begin{figure}
  \centering
  \rotatebox{0}{\scalebox{.35}{\includegraphics{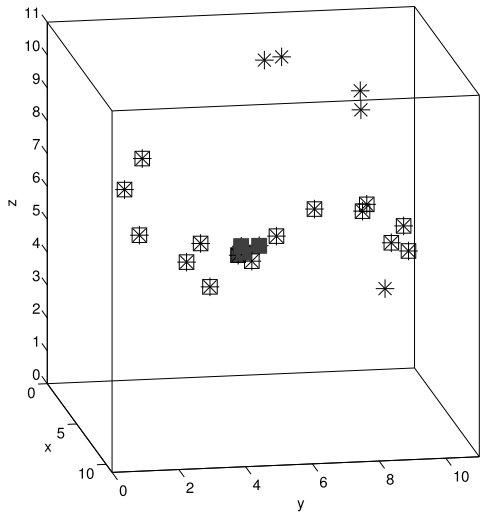}}}
  \caption{ The galaxy distribution within 5 $h^{-1} \, {\rm Mpc}$ of
    one of the SDSS DR7 galaxies that was identified to have $D
    \approx 1$ across the length-scales 0.5 to 5 $h^{-1} {\rm
      Mpc}$. The galaxies are all shown with crosses.  We have used a
    Friend-of-Friend (FoF) algorithm with a linking length of
    $\sqrt{3} \, h^{-1} \, {\rm Mpc}$ to highlight any connected
    structure in the galaxy distribution in this figure.  This yields
    a single filamentary structure containing the majority of the
    galaxies. All the galaxies belonging to this connected structure
    are shown with cells. The filled cells represents galaxies with $D
    \approx 1$. It was not possible to determine $D$ for the other
    galaxies in this figure. }
\label{fig:new1}
\end{figure}

\begin{figure}
  \centering
  \rotatebox{0}{\scalebox{.35}{\includegraphics{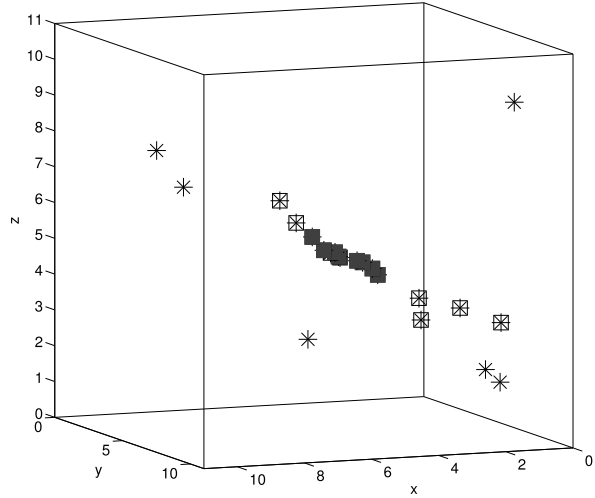}}}
  \caption{Same as Figure  \ref{fig:new1},  for one of the galaxies
    with $D \approx 1$  from  the Millennium Data.} 
\label{fig:new2}
\end{figure}

In order to illustrate our method of classifying cosmological
structures we show the galaxy distribution in the vicinity of one of
the SDSS galaxies that was identified to have $D \approx 1$ across the
length-scales 0.5 to 5 $h^{-1} {\rm Mpc}$ (Figure \ref{fig:new1}).
The expected filament is clearly visible passing through the center of
the figure. We note that there are several galaxies with $D=1$
arranged along the filament visible in the figure.  Figure
\ref{fig:new2} shows a similar plot for one of the galaxies with $D
\approx 1$ from the Millennium data.

\section{Results and Conclusions}

\begin{figure}
\centering
%\rotatebox{-90}{\includegraphics[height=0.4\textwidth]{data_D_plot.ps}}
%\rotatebox{-90}{\scalebox{.35}{\includegraphics{data_D_plot.ps}}}
\rotatebox{-90}{\scalebox{.35}{\includegraphics{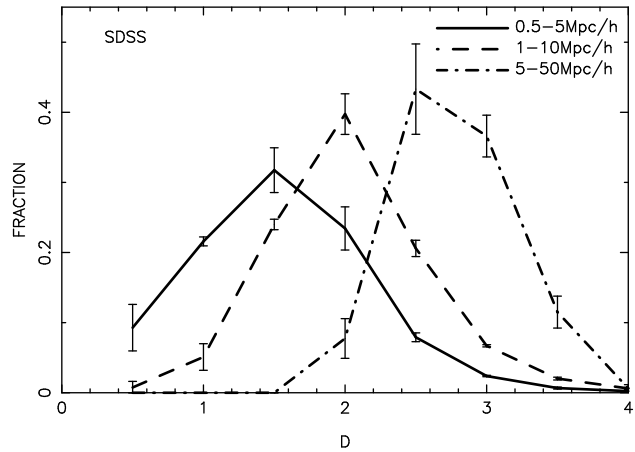}}}
\caption{This shows the fraction of galaxies with a particular $D$
  value for the SDSS data.  The three different curves correspond to
  $D$ values that were determined using the length-scales $0.5-5 \,
  h^{-1} \, {\rm Mpc}$, $1-10 \, h^{-1} \, {\rm Mpc}$ and $5-50 \,
  h^{-1} \, {\rm Mpc}$ respectively. The bins in $D$ have size $\pm
  0.25$.}
  \label{fig:1}
\end{figure}

\subsection{Distribution of $D$ values}

We first analyse the fraction of centers with different $D$ values,
shown in Figure \ref{fig:1}.  The $D$ values were divided into bins of
width $\pm 0.25$. The error bars in the data have been estimated using
bootstrap re-sampling of the data.  Ten bootstrap samples were used
for this purpose.

The solid curve in Figure \ref{fig:1}, which corresponds to the
results for 0.5 to 5 $h^{-1} {\rm Mpc}$, shows a broad peak with a
maxima at $D=1.5$. The bin centered at $D=2$ contains the second
largest fraction of galaxies. These two bins together contain more
than $50\%$ of the centers for which $D$ could be determined.  This
value indicates that the galaxies in the Cosmic Web are predominantly
contained in sheets and filaments at the length-scales 0.5 to 5
$h^{-1} {\rm Mpc}$, with the sheets being somewhat more dominant than
the filaments.  The dashed curve in Figure \ref{fig:1}, which
corresponds to length-scales 1 to 10 $h^{-1} {\rm Mpc}$, shows a sharp
peak at $D=2$. The fraction of galaxies in the neighbouring bins
($D=1.5$ and $2.5$) falls to nearly half the fraction in this bin. The
three bins at $D=1.5,\, 2$ and $2.5$ together contains more than
$80\%$ of the centers for which $D$ could be determined. This
indicates that the galaxies in the Cosmic Web are predominantly in
sheets over the length-scale 1 to 10 $h^{-1} {\rm Mpc}$. The
dot-dashed curve in Figure \ref{fig:1}, which corresponds to the
length-scales 5 to 50 $h^{-1} {\rm Mpc}$, peaks at $D=2.5$ and
$D=3$. The $D=2.5$ contains the maximum no of centers. The two bins
combinely contains more than $70\%$ of the centers. This indicates
that the galaxies in the Cosmic Web are predominantly in sheets and
clusters (volume filling structures) over the range of length-scales 5
to 50 $h^{-1} {\rm Mpc}$. Further, it is interesting to note that we
do not find any center with $D=1$ or $1.5$ at this range of
length-scales. This indicates the complete absence of filamentary
structures at the largest range of length-scales (5 to 50 $h^{-1} {\rm
  Mpc}$) that we have probed.

There is a shift to larger $D$ values in Figure \ref{fig:1} as we
progressively consider larger length-scales. The entire curve showing
the fraction of centers as a function of $D$ shifts to the right when
we consider larger length-scales.  It is quite evident from this that
the nature of the structural elements that make up the Cosmic Web
differs depending on the length-scale at which we view the Cosmic Web.
At small scales $(0.5-5 \, h^{-1} {\rm Mpc})$ we have a mixture of
sheets and filaments. The fraction of sheets increases at larger
length-scales $(1-10 \, h^{-1} {\rm Mpc})$, whereas we predominantly
have a combination of sheets and clusters at the largest length-scale
$(5-50 \, h^{-1} {\rm Mpc})$ that we have probed.  Filaments are
completely absent at the largest length-scale.

\begin{figure}\
\centering
%\rotatebox{-90}{\includegraphics[height=0.4\textwidth]{mill_D_plot.ps}}
%\rotatebox{-90}{\scalebox{.35}{\includegraphics{mill_D_plot.ps}}}
\rotatebox{-90}{\scalebox{.35}{\includegraphics{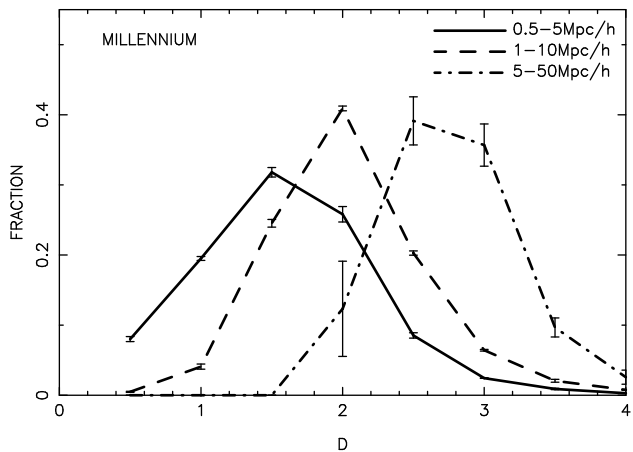}}}
\caption{This shows the fraction of galaxies with a particular $D$
  value for the data from the Millennium simulation.  The three
  different curves correspond to $D$ values that were determined using
  the length-scales $0.5-5 \, h^{-1} \, {\rm Mpc}$, $1-10 \, h^{-1} \,
  {\rm Mpc}$ and $5-50 \, h^{-1} \, {\rm Mpc}$ respectively. The bins
  in $D$ have size $\pm 0.25$.}
  \label{fig:2}
\end{figure}

For comparison, we have also applied the Local Dimension to analyse
the galaxy distribution in a semi analytic galaxy catalogue from the
Millennium Simulation.  We have used the semi analytic galaxy
catalogue \citep{guo11} to extract three different data samples with exactly the
same geometry and galaxy number density as our SDSS sample.  These
three simulated data samples were analysed in exactly the same way as
the actual data.  The results showing the fraction of centers at
different $D$ values are presented in Figure \ref{fig:2}.  The three
different curves in this figure correspond to the same range of
length-scales as three different curves in Figure \ref{fig:1}.  We
find that the fraction of centers with different $D$ values have very
similar distributions in the actual SDSS data and the Millennium
Simulation. These curves for the SDSS data are nearly identical to
those from the Millennium Simulation except for the fact that at
largest length-scale ($5-50 \, h^{-1} \, {\rm Mpc}$), the fraction of
centers with $D=2.5$ is larger in the SDSS data as compared with the
Millennium simulation. This difference in the fraction of centers lies
in the $1-\sigma$ error-bar.

\subsection{Environment dependence}

\begin{figure}
\centering
\rotatebox{-90}{\scalebox{.35}{\includegraphics{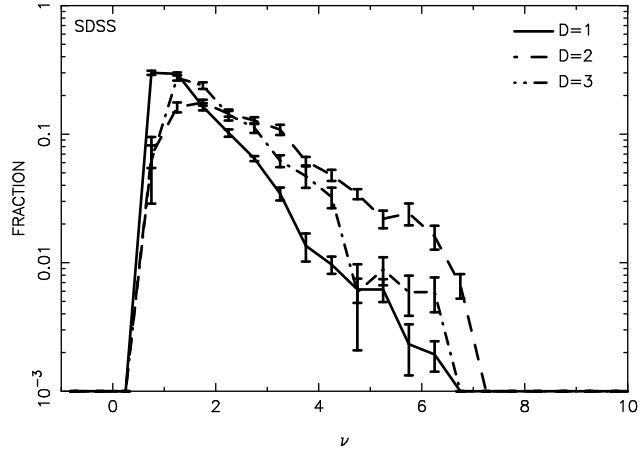}}}
\caption{ The three curves which correspond to $D=1,2$ and $3$
  respectively show the fraction of galaxies as a function of $\nu$.
  The results are for the SDSS data using $R_1= 0.5 \, h^{-1} \, {\rm
    Mpc}$, $R_2=5 \, h^{-1} \, {\rm Mpc}$ and $R_s=1.58 \, h^{-1} \,
  {\rm Mpc}$. }
\label{fig:3}
\end{figure}

\begin{figure}
\centering
\rotatebox{-90}{\scalebox{.35}{\includegraphics{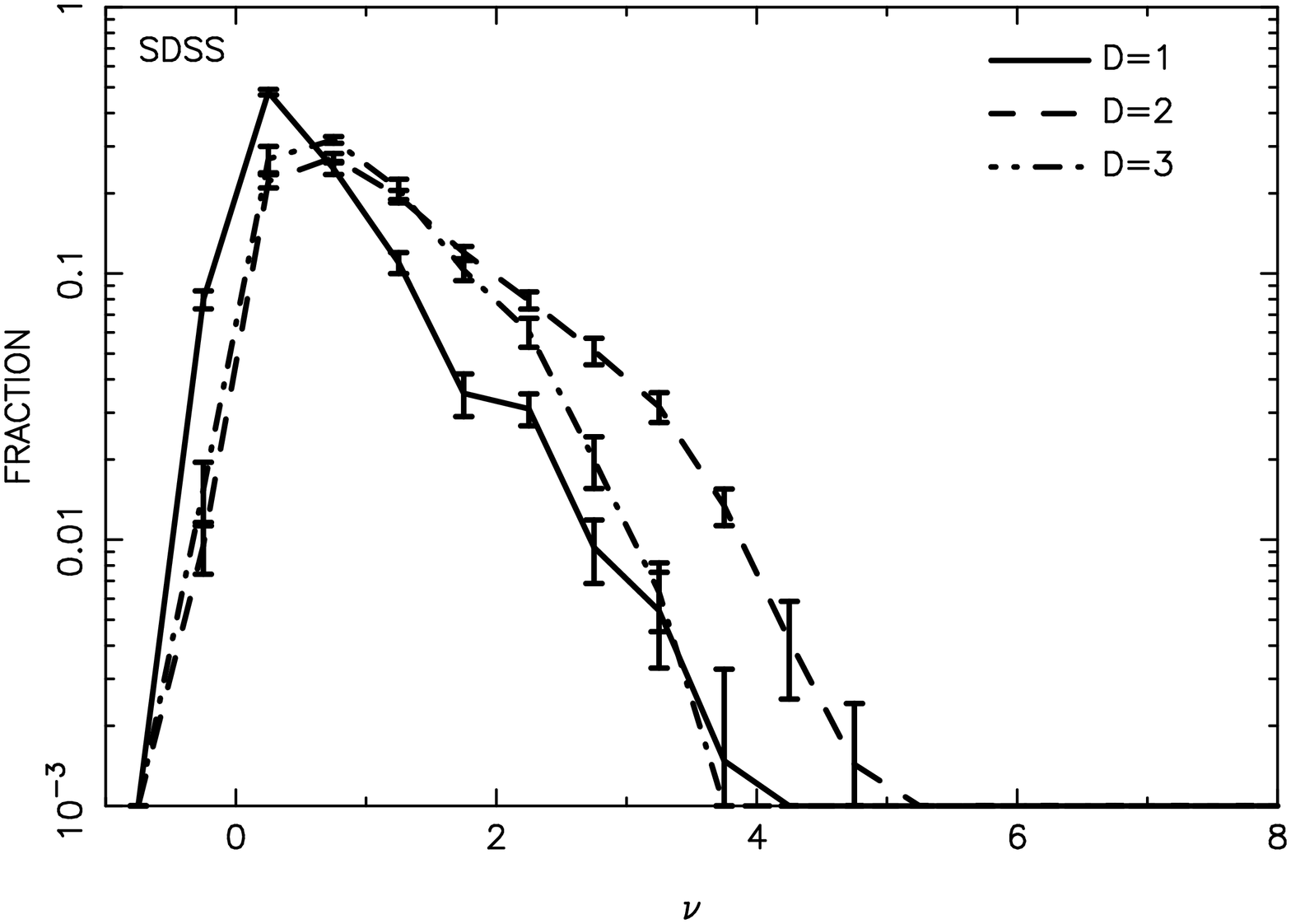}}}
\caption{ The three curves which correspond to $D=1,2$ and $3$
  respectively show the fraction of galaxies as a function of $\nu$.
  The results are for the SDSS data using $R_1= 1 \, h^{-1} \, {\rm
    Mpc}$, $R_2=10 \, h^{-1} \, {\rm Mpc}$ and $R_s=3.16 \, h^{-1} \,
  {\rm Mpc}$. }
\label{fig:4}
\end{figure}

\begin{figure}
\centering
\rotatebox{-90}{\scalebox{.35}{\includegraphics{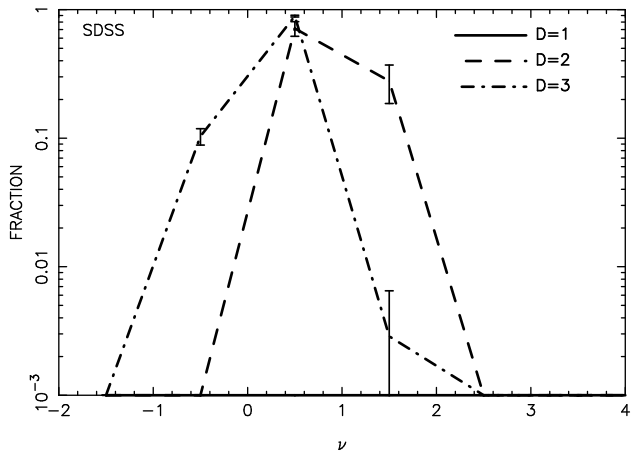}}}
\caption{ The three curves which correspond to $D=1,2$ and $3$
  respectively show the fraction of galaxies as a function of $\nu$.
  The results are for the SDSS data using $R_1= 5 \, h^{-1} \, {\rm
    Mpc}$, $R_2=50 \, h^{-1} \, {\rm Mpc}$ and $R_s=15 \, h^{-1} \,
  {\rm Mpc}$. }
\label{fig:4a}
\end{figure}

In the previous subsection, we have seen that the nature of the
structural elements that make up the Cosmic Web changes depending on
the length-scale at which we view the web.  Our investigations show
that there is a progressive transition from filaments to sheets and
then clusters as we go from smaller to larger scales. In this
subsection, for a fixed range of length-scales, we investigate if the
distribution of a particular kind of structural element is related to
the density environment. As mentioned earlier, the nature of the
structural elements ({\it ie.} $D$ value) depends on the range of
length-scale used to determine $D$. The density environment too
depends on the length-scale at which we smooth the density field.  The
length-scale range $R_1$ to $R_2$ used for estimating $D$ should thus
be consistent with $R_s$, the smoothing length-scale for the density
field.  In our analysis we have chosen $R_s$ to be the geometric mean
of $R_1$ and $R_2$ {\it i.e} $R_s=\sqrt{R_1 R_2}$.  The geometric mean
was chosen instead of the algebraic mean because the latter is
expected to be more biased towards $R_2$. We expect the geometric mean
to give a more representative estimate of the range of length-scales
$R_1$ to $R_2$ which span a decade.

For fixed values of $R_1$, $R_2$ and $R_s$ we focus on the
distribution of the centers with a particular $D$ value. For this
purpose. the $D$ values were divided into bins of width $\pm 0.5$
centered at $D=1,\,2$ and $3$.  To determine the density environment,
we have converted the entire galaxy distribution to a density field on
a grid of spacing $[0.5 \, h^{-1} {\rm Mpc}]^3$ using the
Cloud-in-Cell method. This density field is then smoothed with a
Gaussian kernel having a smoothing length $R_S$.  The smoothing is
carried out in Fourier space by multiplying the Fourier transform of
the density field with $\exp(-k^2 R_s^2/2)$ and transforming back to
real space.  The density field at any grid point was quantified using
the dimensionless ratio $\nu=\delta/\sigma$ where $\delta=\delta
\rho/\bar\rho$ is the density contrast of the smoothed density field
at the particular grid point and $\sigma$ is the standard deviation of
the density contrast of the smoothed density field evaluated using all
the grid points that lie within the survey volume.  Considering only
the centers for which it is possible to determine a $D$ value, we use
the value of $\nu$ at the grid point nearest to the center to assign a
$\nu$ value to each of these centers. The value of $\nu$ associated
with any of the galaxies gives an estimate of the density environment
in the vicinity of the structural element centered on that galaxy.

We first consider the range, $R_1= 0.5 \, h^{-1} \,{\rm Mpc}$ and
$R_2=5 \, h^{-1} \, {\rm Mpc}$ for which $R_s = 1.58 \, h^{-1} \, {\rm
  Mpc}$ and $\sigma = 5.15$.  Figure \ref{fig:3} shows the results for
this range of length-scales.  We first consider only the centers with
$D=1$ for which the results are shown in the solid curve of this
figure. This curve shows the fraction of centers with a particular
$\nu$ value. The $\nu$ values were divided into bins spanning $\pm
0.25$ for this purpose.  The dashed and dash-dotted curves
respectively show the corresponding results for $D=2$ and $3$
respectively. We find that the distribution of the fraction of centers
as a function of $\nu$ is qualitatively similar for all three values
of $D$.  The fraction shows a peak near $\nu \sim 1$, with a very
sharp decline in the fraction at $\nu <1$ and a relatively gradual
decline at $\nu>1$.  While the behaviour is qualitatively similar,
there are quantitative differences between the different $D$ values.
We see that the fraction peaks at a somewhat smaller $\nu$ value for
$D=1$ as compared to $D=2$ and $3$.  At $\nu <1$ the values of the
fraction are somewhat larger for $D=1$ as compared to $D=2$ and $D=3$.
Further, the values of the fraction in the vicinity of the peak are
somewhat larger for $D=2$ as compared to $D=3$.  The behaviour is
reversed for $\nu>1$.  The fraction is smaller for $D =1$ in
comparison to $D=2$ and $3$ in the range $1 < \nu < 3$. For $\nu >3$,
we find that $D=3$ is roughly consistent with $D=1$, whereas the
fraction is considerably higher for $D=2$.

Figure \ref{fig:4} shows the same quantities as Figure \ref{fig:3}
with the difference that the range of length-scales now corresponds to
$R_1=1$ and $R_2=10$ $h^{-1}\,{\rm Mpc}$ for which $R_s= 3.16\, h^{-1}
\, {\rm Mpc}$ and $\sigma= 3.09$.  The behaviour, we find, is very
similar to that in Figure \ref{fig:3} except that the peak has shifted
to a smaller $\nu$ value ($\nu \sim 0.5$).  For $\nu <0.5$, the
fraction is larger for $D=1$ in comparison to $D=2$ and $3$. This is
reversed for $\nu>0.5$ where the fraction is smaller for $D=1$ in
comparison to $D=2$ and $3$.  There is another transition around $\nu
\sim 2$, where for $D=3$ the fraction falls below that of $D=2$.  We
find that, for $\nu>2$, $D=1$ and $3$ are comparable and are lower
than $D=2$.

Figure \ref{fig:4a} shows the result for $R_1=5$ and $R_2=50 \, h^{-1}
\, {\rm Mpc}$ for which $R_s = 15.8 \, h^{-1} \, {\rm Mpc}$ and $\sigma
= 0.97$.  The $\nu$ values were divided into bins spanning $\pm 0.5$
for this purpose. The behaviour is different compared to that in
Figure \ref{fig:3} and Figure \ref{fig:4}.  The analyse for this
length-scale shows total absence of center in a filament.  For $\nu <
1.5$, the fraction for $D=3$ larger in comparison to $D=2$. This
situation is reverse for $\nu > 1.5$, where the fraction is smaller
for $D=2$ in comparison to $D=3$.

\begin{figure}
\centering
\rotatebox{-90}{\scalebox{.35}{\includegraphics{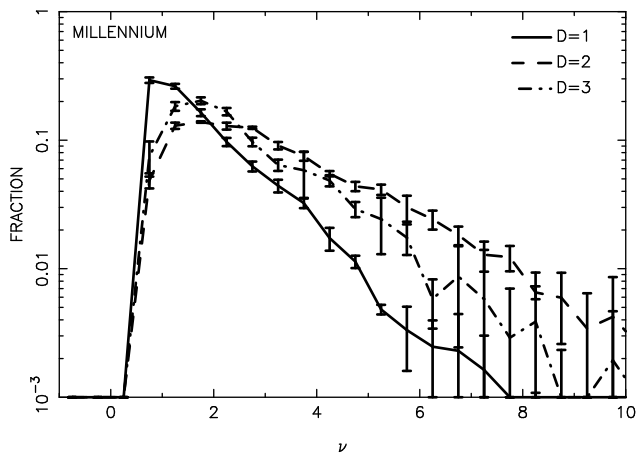}}}
\caption{ The three curves which correspond to $D=1,2$ and $3$
  respectively show the fraction of galaxies as a function of $\nu$.
  The results are for the data from the Millennium Simulation using
  $R_1= 0.5 \, h^{-1} \, {\rm Mpc}$, $R_2=5 \, h^{-1} \, {\rm Mpc}$
  and $R_s=1.58 \, h^{-1} \, {\rm Mpc}$. }
\label{fig:6}
\end{figure}
\begin{figure}
\centering
\rotatebox{-90}{\scalebox{.35}{\includegraphics{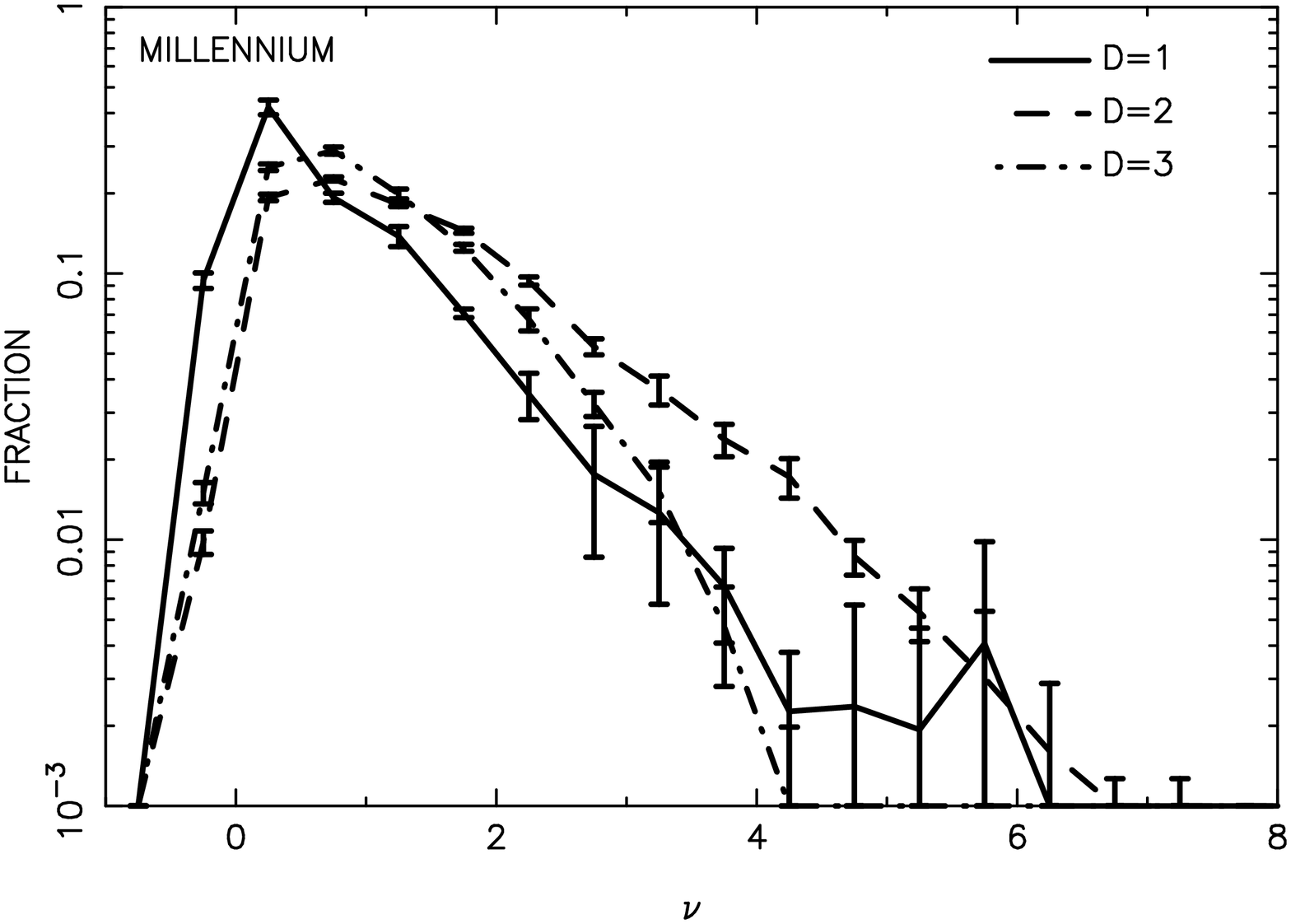}}}
\caption{ The three curves which correspond to $D=1,2$ and $3$
  respectively show the fraction of galaxies as a function of $\nu$.
  The results are for the data from the Millennium Simulation using
  $R_1= 1 \, h^{-1} \, {\rm Mpc}$, $R_2=10 \, h^{-1} \, {\rm Mpc}$ and
  $R_s=3.16 \, h^{-1} \, {\rm Mpc}$. }
\label{fig:7}
\end{figure}
\begin{figure}
\centering
\rotatebox{-90}{\scalebox{.35}{\includegraphics{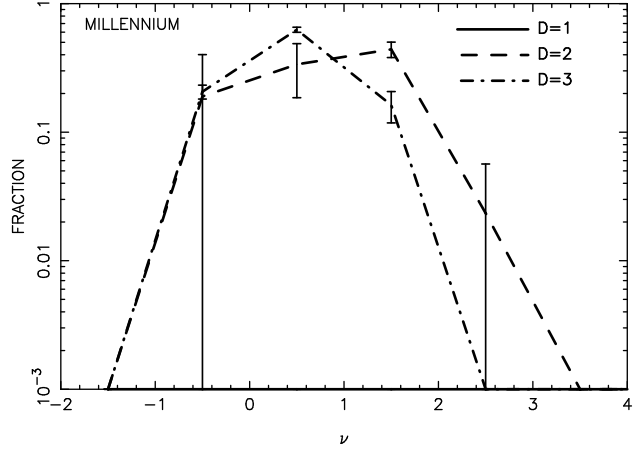}}}
\caption{ The three curves which correspond to $D=1,2$ and $3$
  respectively show the fraction of galaxies as a function of $\nu$.
  The results are for the data from the Millennium Simulation using
  $R_1= 5 \, h^{-1} \, {\rm Mpc}$, $R_2=50 \, h^{-1} \, {\rm Mpc}$ and
  $R_s=15.8 \, h^{-1} \, {\rm Mpc}$. }
\label{fig:7a}
\end{figure}

We may interpret the different curves in Figures \ref{fig:3} and
\ref{fig:4} as representing the probability of finding a particular
kind of structural element in the density environment corresponding to
$\nu$. For example, the curve for D=1 in Figure \ref{fig:3} gives the
probability of finding a filament in the interval $\nu-0.25$ to
$\nu+0.25$ for different values of $\nu$. Similarly, the curves for
D=2 and D=3 show the probability of finding a sheet and a cluster
respectively.  Our analysis shows that the filaments, sheets and
clusters have different probability distributions. The filaments are
preferentially distributed in low density environments relative to the
distribution of sheets and clusters.  We have a cross-over from this
behaviour at intermediate densities.  The sheets are preferentially
distributed relative to filaments and clusters in the high density
environments.  The $\nu$ value where these transitions occur depends
on our choice of $R_1$ and $R_2$. For 0.5 to 5 $h^{-1}\,{\rm Mpc}$,
the density contrasts ranges are $\nu<1$, $1 < \nu < 3$ and $\nu>3$
while for 1 to 10 $h^{-1}\,{\rm Mpc}$ the $\nu$ ranges are $\nu< 0.5$,
$0.5 < \nu < 2$ and $\nu>2$.  These findings indicate that the way in
which different structural elements are distributed along the Cosmic
Web depends jointly on two factors (a) the local density environment,
and (b) the length-scale at which we view the Cosmic Web.

For comparison, the above analysis was also performed using the galaxy
distribution in the Millennium Simulation. The analysis was carried
out in exactly the same way as for the actual SDSS DR7 data. The
results for the Millennium simulation shown in Figures \ref{fig:6},
\ref{fig:7} and \ref{fig:7a} are analogous to the plots shown for the
SDSS DR7 in Figures \ref{fig:3}, \ref{fig:4} and \ref{fig:4a}
respectively.  We find that the results from the Millennium Simulation
are very similar to those obtained for the actual SDSS DR7 for
length-scale 0.5 to 5 $h^{-1} {\rm Mpc}$ and 1 to 10 $h^{-1} {\rm
  Mpc}$. The results for the length-scale 5 to 50 $h^{-1} {\rm Mpc}$
from Millennium Simulation differs to those obtained from the SDSS
DR7.  The $\nu$ values, in  Figure \ref{fig:7a}, were divided into
bins spanning $\pm 0.5$.  At $\nu <-0.5$ the values of the fraction
are somewhat similar  for $D=2$ and $D=3$. The fraction is smaller for
$D=2$ as compared to $D=3$ in the range $-0.5 < \nu < 1$. This trend
reversed for $\nu > 1$, where $D=2$ is larger compared to $D=3$.

In summary the analysis of both the SDSS DR7 and the Millennium
Simulation exhibit similar trends. The Local Dimensions were estimated
separately in three different ranges of length-scales, 0.5 to 5
$h^{-1} {\rm Mpc}$, 1 to 10 $h^{-1} {\rm Mpc}$ and 5 to 50 $h^{-1}
{\rm Mpc}$. We find that there is a progressive shift in the $D$
values as we move to larger length-scales. At small length-scales
there is a mixtures of sheets and filaments, and the fraction of
sheets increases as we move to larger length-scales (1 to 10 $h^{-1}
{\rm Mpc}$). We find that sheets and clusters are the predominant
structures at the largest length -scales (5 to 50 $h^{-1} {\rm Mpc}$).
Filaments are absent at this length-scales. It is interesting to note
that \citet{2009MNRAS.396.1815F} find that the mass fraction of
filaments decreases while that of sheet increases with an increase in
the smoothing length-scale, which is consistent with the results of
this paper.

The gradual transition, with increasing length-scales, from filaments
$(D=1)$ to sheets ($D=2$) and then to clusters ($D=3$) is indicative
of and consistent with a transition to homogeneity ($D=3$) at large
length-scales.  A recent analysis of the SDSS Main galaxy sample
\citep{sarkar} shows that there is transition to homogeneity at
$60-70\,h^{-1} \,{\rm Mpc}$, and that the galaxy distribution is
consistent with a homogeneous point distribution at length-scales
larger than this. An earlier study \citep{pandey} had analysed thin,
nearly 2-D, sections through the galaxy distribution and had found
evidence for statistically significant filamentary patterns to
length-scales as large as 80 $h^{-1}\,{\rm Mpc}$. This is apparently
inconsistent with the findings of the present paper which fails to
find any filaments that span across the length-scale 5 to 50
$h^{-1}\,{\rm Mpc}$.  It should however be noted that the structures
which were identified as filaments in the 2-D sections are quite
likely to be sheets when the structures are viewed in full 3-D.

\section*{Acknowledgment}

PS would like to acknowledge Senior Research Fellowship of the
University Grants Commission (UGC), India, for providing financial
support during which a part of this work was done. BP acknowledges the
Center for Theoretical Studies (CTS), Indian Institute of Technology,
Kharagpur, for providing support to visit CTS. BP would also like to
thank the Alexander von Humboldt Foundation for support through a
post-doctoral fellowship.

The Millennium Simulation data bases \citep{2006astro.ph..8019L} used
in this paper and the web application providing online access to them
were constructed as part of the activities of the German Astrophysical
Virtual Observatory.

The SDSS DR7 data were downloaded from the SDSS skyserver
http://cas.sdss.org/dr7/en/. Funding for the creation and distribution
of the SDSS archive has been provided by the Alfred P. Sloan
Foundation, the Participating Institutions, the National Aeronautics
and Space Administration, the National Science Foundation, the US
Department of Energy, the JapaneseMonbukagakusho and the Max
Planck Society. The SDSS web site is http://www.sdss.org/.

The SDSS is managed by the Astrophysical Research Consortium (ARC) for
the Participating Institutions. The Participating Institutions are The
University of Chicago, Fermilab, the Institute for Advanced Study, the
Japan Participation Group, The Johns Hopkins University, the Korean
Scientist Group, Los Alamos National Laboratory,
theMax-Planck-Institute forAstronomy (MPIA), the Max-Planck-Institute
for Astrophysics (MPA), New Mexico State University, University of
Pittsburgh, Princeton University, the United States Naval Observatory
and the University ofWashington.

\end{document}